\documentclass[12pt,preprint]{aastex}
\usepackage{graphicx}

\newcommand{\La}{Ly$\alpha$}
\newcommand{\Ot}{[O\,{\sc iii}]}
\newcommand{\Od}{[O\,{\sc ii}]}
\newcommand{\Ha}{H$\alpha$}

\begin{document}

\title{Search for z$\sim$7 \La\ emitters with Suprime-Cam at the Subaru Telescope. }
\author{P.Hibon\altaffilmark{1,2}, 
N.Kashikawa\altaffilmark{3,4}, 
C.Willott\altaffilmark{5}, 
M.Iye\altaffilmark{3,4}, 
T.Shibuya\altaffilmark{3,4}
}

\begin{abstract}
We report a search for z=7 \La\ emitters (LAEs) using a custom-made Narrow-Band filter,
centered at 9755\AA\, with the instrument Suprime-Cam installed at the
Subaru telescope. 
We observed two different fields and obtained two
sample of 7 \La\ emitters of which 4 are robust in each field. We are covering the luminosity range of
$9.10^{42} - 2.10^{43}\mathrm{erg}\, \mathrm{s}^{-1}$ in comoving volumes of $\sim 4\times10^{5} \textrm{ and } 4.3\times10^{5} Mpc^{3}$.\\
From this result, we derived possible z$\sim$7 \La\ luminosity functions for the full samples and for a subsample of 4 objects in each field. We do not observe, in each case, any strong evolution between the z=6.5 and z$\sim$7 \La\ luminosity functions. 
Spectroscopic confirmation for these candidate samples is required to establish a definitive measure of the luminosity function at z$\sim$7.
\end{abstract}

\altaffiltext{1}{Gemini Observatory, La Serena, Chile;
email phibon@gemini.edu}
\altaffiltext{2}{School of Earth and Space Exploration,  Arizona  State University,  Tempe, AZ  85287}
\altaffiltext{3}{Optical and Infrared Astronomy Division, National Astronomical Observatory, Mitaka, Tokyo 181-8588, Japan}
\altaffiltext{4}{Department of Astronomy, School of Science, Graduate University for Advanced Studies, Mitaka, Tokyo 181-8588, Japan}
\altaffiltext{5}{Herzberg Institute of Astrophysics, National Research Council,5071 West Saanich Road, Victoria, BC V9E 2E7, Canada}





\section{Introduction}

Galaxies formed at high redshifts play a key role in understanding how and when the reionization of the universe took place.  They also help constrain the physical mechanisms that drove the formation of the first stars and galaxies in the universe.  
Over the last decade, the limits of the observable universe have been pushed at z $\sim 6$, which corresponds to $\sim 90\%$ the age of the universe. Their detection will allow us to probe the era of reionization. \\ 
 Selecting galaxies with strong
emission lines, such as \La\ line,  allows us to probe the high-redshift \La\ luminosity function.\\
The search for the redshifted \La\ emission at the longest possible wavelength is complicated by the presence of OH emission lines within the terrestrial atmosphere. This strong emission line is responsible for the faintness limit at which celestial objects can be detected with ground-based telescopes at near-infrared (IR) wavelengths.  Fortunately, there are spectral intervals with lower OH-background that allow for a fainter detection limit from the ground-based observations.  This is known as the narrow-band (NB) imaging technique.  \\
This is one of the most successful methods to detect strong Ly$\alpha$ emission lines of galaxies, since it relies on a specific redshift interval as well as a selected low-sky background window.  This filter allows us for maximum detection of light from the Ly$\alpha$ emitters at the central wavelength, while minimizing the adverse influences of sky emission.  \\
	  Over hundred of $z > 6$ LAEs have been photometrically selected and spectroscopically identified in this way since \cite{Hu2002}.  
\cite{Kashikawa2011} obtained an exhaustive sample of LAEs, (45 at z=6.5 and 54 at z=5.7 spectroscopically confirmed), from which a first robust estimate of the \La\ LF was derived. This LF shows an apparent deficit compared to the z=5.7 \La\ LF of \cite{Shimasaku2006}, corresponding to a possible luminosity evolution from z=5.7 to z=6.5 of $L^{*}_{z=6.6} \sim 0.4-0.6 L^{*}_{z=5.7}$. They conclude that the reionization of the universe is not been completed at 6.5.\\
\cite{Ouchi2010} have now obtained the largest sample to date of 207 LAEs at z=6.6 with the NB imaging technique. Their derived z=6.6 apparent \La\ LF indicate a decrease from 5.7 at 90\% confidence level with a more dominant decrease of luminosity evolution ($L^{*}$) than number evolution ($\Phi^{*}$), in agreement with \cite{Kashikawa2006}. They claim therefore that the hydrogen in the IGM is not highly neutral at z=6.6.\\
\cite{Hu2010} have obtained 88 z$\sim$5.7 and 30 z$\sim$6.5 \La\ emitters. Their results on the evolution of the \La\ LF are in agreement with previous works from \cite{Malhotra2004} and \cite{Kashikawa2006}.\\
\cite{Iye2006} have first confirmed spectroscopically a z=6.96 LAE. From this result, \cite{Ota2008} assumed an evolution of density from z=5.7 to z$\sim$7. They found that the IGM is not highly neutral at z$\sim$7 and the neutral hydrogen fraction could evolve from $x_{HI}^{z=6.6} \sim 0.24-0.36$ to $x_{HI}^{z\sim 7} \sim 0.32-0.64$. 
\cite{Hibon2010} has also performed a search for z$\sim$7 LAEs using the IMACS instrument at the Magellan telescope. This study resulted in a sample of 6 z$\sim$7 LAEs candidates, for which the spectroscopic follow-up data are in analysis.\\


We present here a new NB imaging survey with the Suprime-Cam/Subaru telescope -- targeting $z\sim 7$ Lyman-$\alpha$ emitters with the custom-made $NB973$ filter ($\lambda_{\mathrm{center}}$=9755\AA, FWHM=200\AA).  This paper first presents the data (Section 1) and the data reduction procedure (Section 2).  We then describe the method of selection and contamination of low-redshift interlopers for high redshift LAEs in Section 3.  We present the final sample of $z\sim 7$ LAEs and Ly$\alpha$ luminosity function at this redshift in Section 4. 

Throughout this study, we adopt the following cosmological parameters :
$H_{0}=70km.s^{-1}.Mpc^{-1}$, $\Omega_{m}=0.3$, $\Omega_{\Lambda}=0.7$ \citep{Spergel2007}.
 All magnitudes are AB magnitudes.

\section{Observations and Data Reduction}

\subsection{Observations}
The data were taken with the Suprime-Cam instrument (the Subaru Prime Focus Camera), installed at the
8m Subaru telescope at the National Astronomy Observatory of Japan. This
instrument delivers a mosaic of ten 2048 $\times$ 4096 CCDs, which covers a 34' x 27' field of view with a pixel scale of 0.20''. The new CCD, with a QE two times better than the precedent one at the red end ($\lambda=8000-10400$\AA), was installed.\\
By taking exposures lasting 20 min, we ensured background limited performance. 
We did offset the telescope between each exposure.
 The offset sequence, for each field and night, was randomly
chosen. \\
We targeted two fields (hereafter called D33 and D41) covered by
 the Canada France Hawaii
Telescope Legacy Survey (CFHT-LS) and the WIRCam Deep Survey (WIRDS :
PIs Willott \& Kneib)\footnote{Based on observations obtained with
  WIRCam, a joint project of CFHT,Taiwan, Korea, Canada, France, at
  the Canada-France-Hawaii Telescope (CFHT) which is operated by the
  National Research Council (NRC) of Canada, the Institute National
  des Sciences de l'Univers of the Centre National de la Recherche
  Scientifique of France, and the University of Hawaii. This work is
  based in part on data products produced at TERAPIX, the WIRDS
  (WIRcam Deep Survey) consortium, and the Canadian Astronomy Data
  Centre. This research was supported by a grant from the Agence
  Nationale de la Recherche ANR-07-BLAN-0228}. The total area of the
survey, the two fields considered, is 2340 square arc minutes.

We observed using the custom-made NB filter centered at 9755\AA\, with a wavelength range of $\Delta\lambda$=200\AA\, ($NB973$) during two nights in July 2009. During these observations, the conditions were partially good : we obtained 15 and 19 20-minute exposures during these nights with a seeing varying between 0.5'' and 0.85'' .\\
The total exposure time for each epoch of data is given in Table \ref{table1}.

\subsection{NB Data Reduction}
Using the Data Reduction Software developed for the Subaru Suprime-Cam instrument \citep{Ouchi2004,Yagi2002}, we performed the following steps: a bias subtraction, a flat fielding, a distortion and atmospheric dispersion correction, a sky subtraction, a bad regions masking such as satellite trails and AG probe, the alignment of the individual exposures, and the final co-adding step resulting on a stacked image. After eliminating the low signal-to-noise regions at the edges of the field of view, we obtain an effective area of 1118 $\textrm{arcmin}^{2}$ for D33 field, and 1202 $\textrm{arcmin}^{2}$ for D41 field.\\
We realize the astrometric calibration on the individual images before
the final stacking using the UCAC2 catalog \citep{Zacharias2004}.\\
We need then to adjust the WCS parameters to
obtain a more precise alignment. For this purpose we use the IRAF task \textit{msctpeak} 
to interactively align the catalog stars in our images and 
update their headers.
We finally obtain an astrometry calibration for each individual images
with a precision of rms$\sim \pm0.1$ arcsec in both directions.\\
The photometric calibration of the $u^\ast$, $g'$, $r'$, $i'$, $z'$ broad band CFHT-LS data (described in the next section) is based on the SDSS 
data for stars with 17$<i'<$21 and the
Megacam-SDSS color transformation equations of \cite{Regnault2009}.The precision
obtained in $u^\ast$, $g'$, $r'$, $i'$, $z'$ is between 0.03 and 0.02 mag.
As the $NB973$ filter is included in the $z'$-band filter, we calibrated our $NB973$ stacked images, 
using the AUTO magnitude from the $z'$ band SExtractor catalog.
 We performed the calibration on 1500 non-saturated stars within 16$<z'<$20.
Considering the photometric error on the broad-band calibration, we obtain a
photometric calibration precise at 0.1 magnitude in $NB973$.\\
To estimate the limiting magnitude for the different bands, we used
the script \textit{limitmag.cl} available in the SDFRED package. 
This task constructs a count distribution from random photometric apertures
on the image. Then it fits a Gaussian
profile into this distribution, obtains the sigma number of this
profile and calculates the limiting magnitude of the image.
We report in Table~\ref{table1} the limiting magnitude for each band
in each field.\\

\subsection{Broad Band Data}
Very deep optical imaging data of our observed fields are available through the CFHT-LS. 
For the purpose of this study, we made use of the
T0006 release. These data products are available from the CADC
archive and take form of image stacks in the $u^\ast$, $g'$, $r'$, $i'$, $z'$
filters and of ancillary data such as weight maps, catalogs etc.  The
spectral curves of the filter $u^\ast$, $g'$, $r'$, $i'$, $z'$ are similar to the ones of SDSS filters.
We also have in hand the Near
Infrared deep Imaging data from the WIRCam Deep Survey (WIRDS) (PIs: C.Willott, J-P. Kneib) (Bielby et al. in preparation).\\
These optical data have been calibrated photometrically using the SDSS photometry
 and the NIR data using 2MASS photometry \citep{McCracken2010}.
Considering internal and external photometric error sources, the uncertainty 
on the optical and the NIR data photometry is $\sim$0.03 mag and 
$\sim$0.02 mag, respectively.\\
As the broad-band data and the narrow-band data do not have the same
pixel scale, we resampled the broad-band data using the software
SWARP to obtain optical and NIR images with 0.2 arcsec/pixel.
The alignment in pixels is then verified using IRAF \textit{geomap/geotran}
tasks.\\
Our complete set of data is therefore scaled at 0.2 arcsec/pixel and
is covering an effective area of 1118 and 1202 $\textnormal{arcminutes}^{2}$, for D33 and D41 fields, respectively. \\

\begin{figure}
\centering
\resizebox{\hsize}{!}{\includegraphics{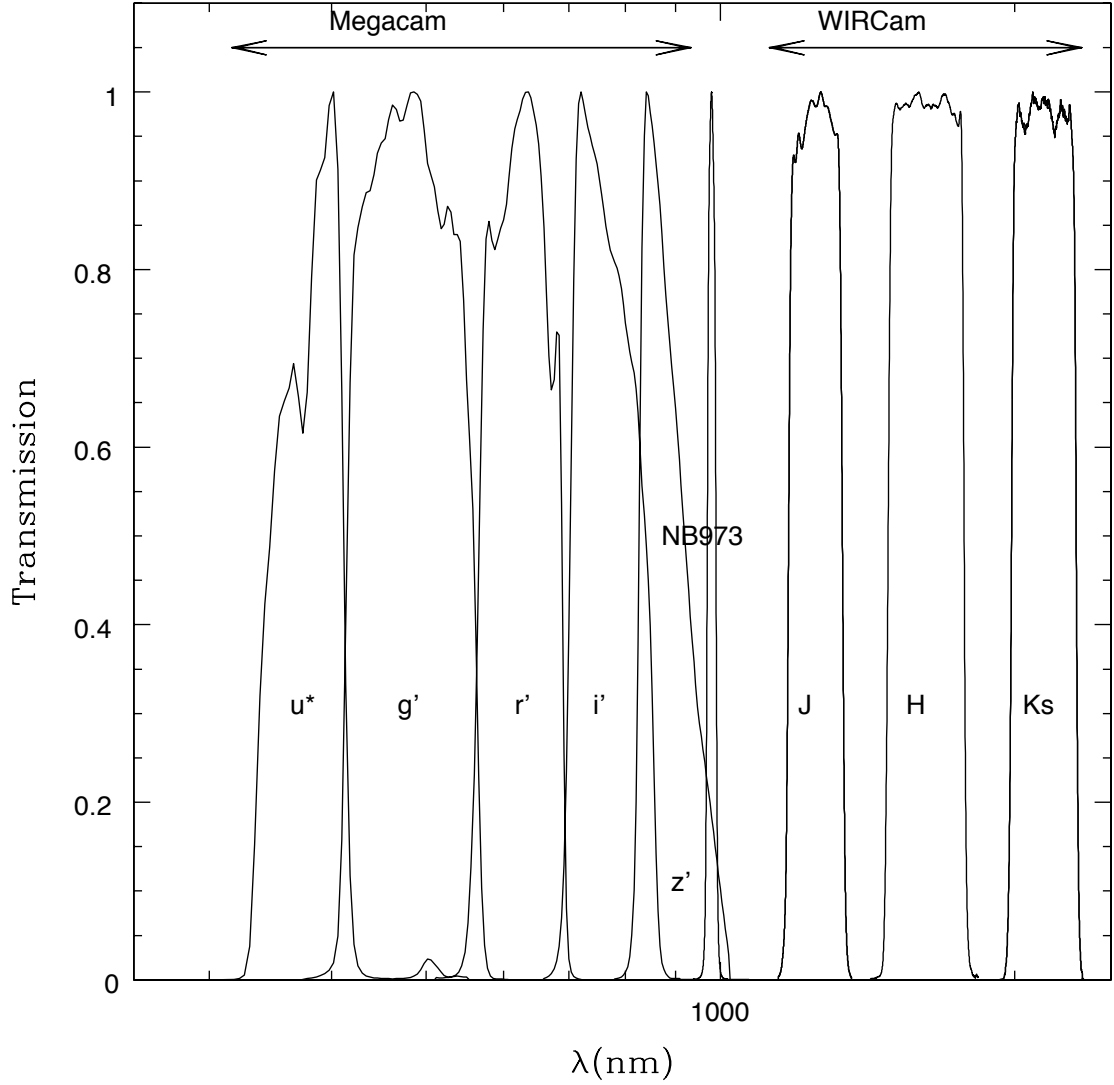}}
\caption{Transmission curves of the filters corresponding to the complete set of data
used in this paper. All transmissions are normalized to 100\% at maximum.}
\label{fig2}
\end{figure}

A summary of the observational data for both fields, D33 and D41, used in this paper is provided in
Table~\ref{table1}.
Figure~\ref{fig2} shows the transmission curves of the filters corresponding
to the multi-band data used in this study. 

\begin{table}
\caption{Observational data.} \label{table1} 
\centering
\begin{tabular}{||l||c ||c |c || c| c||} \hline \hline
 & &\multicolumn{2}{c|}{Integration}& \multicolumn{2}{c|}{Limiting}\\ 
Instrument    & Band  & \multicolumn{2}{c|}{time (hrs)} &  \multicolumn{2}{c|}{magnitudes$^{\mathrm{a}}$}\\
          &               &    D33   &  D41                 &  D33   &  D41 \\ \hline  
MegaCam   &         $u^\ast$    &   21.3      &   21.4      &  27.5     &  27.4\\         
MegaCam   &         $g'$       &    22.1     &   24.4      &  27.5    &  27.2\\         
MegaCam   &         $r'$       &   39.6      &  40.7       &  27.0   &  26.8\\         
MegaCam   &         $i'$       &   69.3      &   65        &  26.6    &  26.5\\         
MegaCam   &         $z'$       &   48.7      &   44.5      &  25.5   &  25.4\\ \hline  
Suprime-Cam       &        $NB973$   &         5   &  6.3        &  24.3   &  24.7 \\ \hline 
WIRCam     &     $J$           &  4.6        & 4.8         &  24.5     &  24.4\\        
WIRCam     &     $H$          &  4.4        & 4.1         &  24.4   &  24.1\\         
WIRCam     & $K_{\mathrm s}$   &  4.8    & 3.9         &  24.2   &  24.3\\ \hline  
\end{tabular}
\begin{list}{}{}
\item[$^{\mathrm{a}}$] $5 \sigma$ magnitude limits in apertures 2\arcsec\
in diameter for MegaCam, WIRCam and Suprime-Cam. 
\end{list}
\end{table}

\section{Sample }
\subsection{Catalog generation}
We generate the catalogs using the software SExtractor of \cite{Bertin1996}. We used the dual-image mode : the first image, the detection image, has been settled as the combined $NB973$ image, the second image, the measurement image, corresponds to the resampled images from the optical and NIR bands.\\
We choose to detect objects in 7 pixels above a threshold of 1.2$\sigma$. The aperture used for the photometry is 1\arcsec.

\subsection{Selection}

\noindent{Criterion\#1 : SNR($NB973_{combined}$)$> 5\sigma$. \\}
We selected objects with a 5$\sigma$ detection on the combined $NB973$ images.\\

\noindent{Criterion\#2 : SNR($u^\ast$, $g'$, $r'$, $i'$, $\chi^{2}$)$< 3\sigma$. \\}
Due to the Gunn- Peterson trough, i.e. the nearly complete absorption
of the flux short-ward of Lyman$\alpha$ as a result of the large neutral
hydrogen column density in the Intergalactic Medium (IGM), we
observe spectral discontinuity at redshifts greater than about 6. We
are therefore searching for objects which are not detectable in
optical ($u^\ast$, $g'$, $r'$, $i'$) bands. A possible method is therefore to select objects with less than a
3$\sigma$ detection in filters blue-ward of the expected \La\ emission :
$u^\ast$, $g'$, $r'$, $i'$. \\
The color break between the optical and $NB973$
filters is high and covers a wide spectral range. 
Moreover, for the
CFHT-LS, the Terapix data center generated deep $\chi^2$ image
combining the $g'$, $r'$ and $i'$ images. We consider therefore this $\chi^2$ image as well as the other blue-ward optical filters and we select objects with less than a 3$\sigma$ detection in the $\chi^2$ image.\\
In summary, we selected objects with less than a 3$\sigma$ detection in $u^\ast$, $g'$, $r'$, $i'$,and $\chi^{2}$ bands.\\

\begin{figure}
\centering
\resizebox{8cm}{!}{\includegraphics{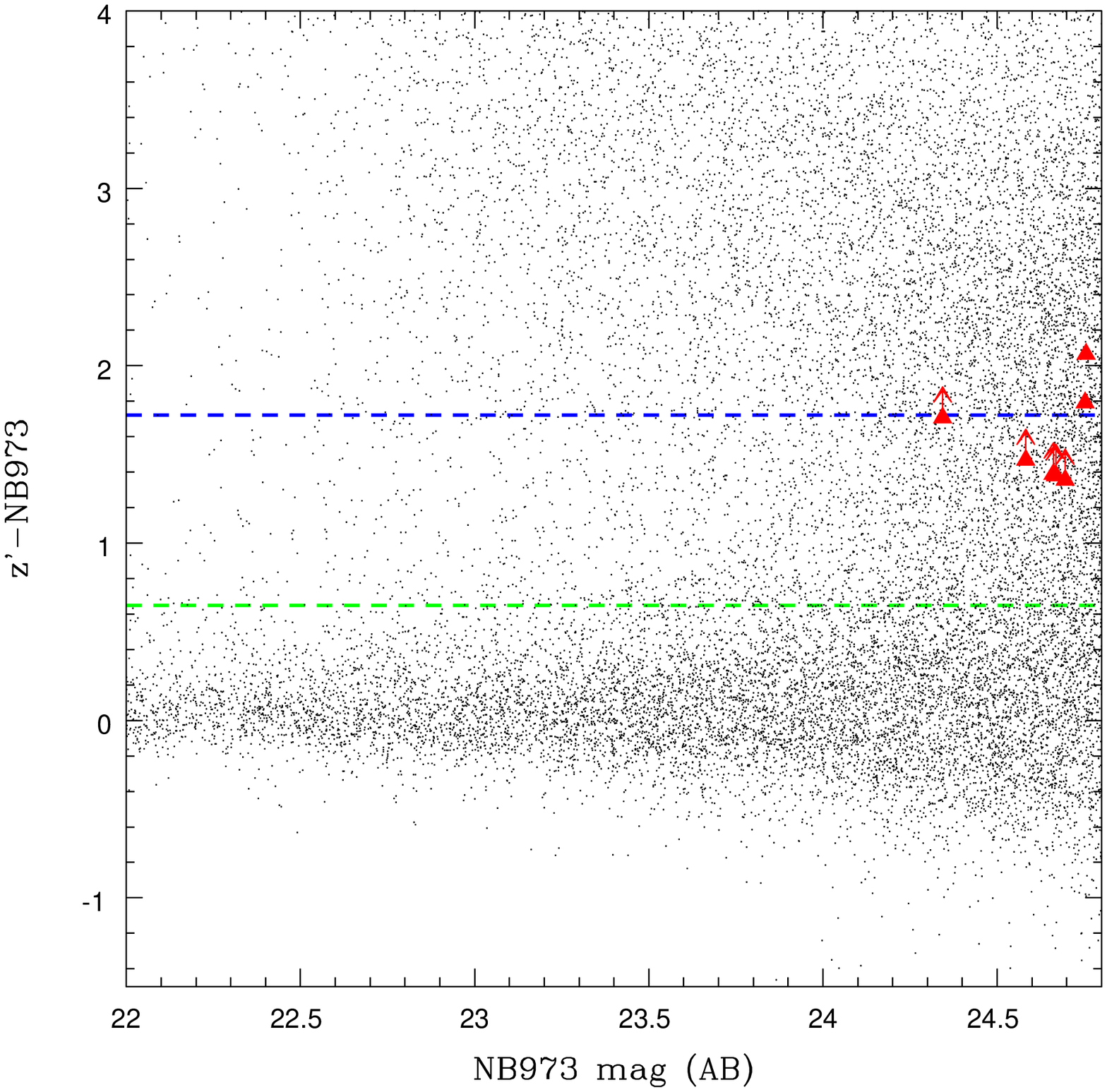}}
\resizebox{8cm}{!}{\includegraphics{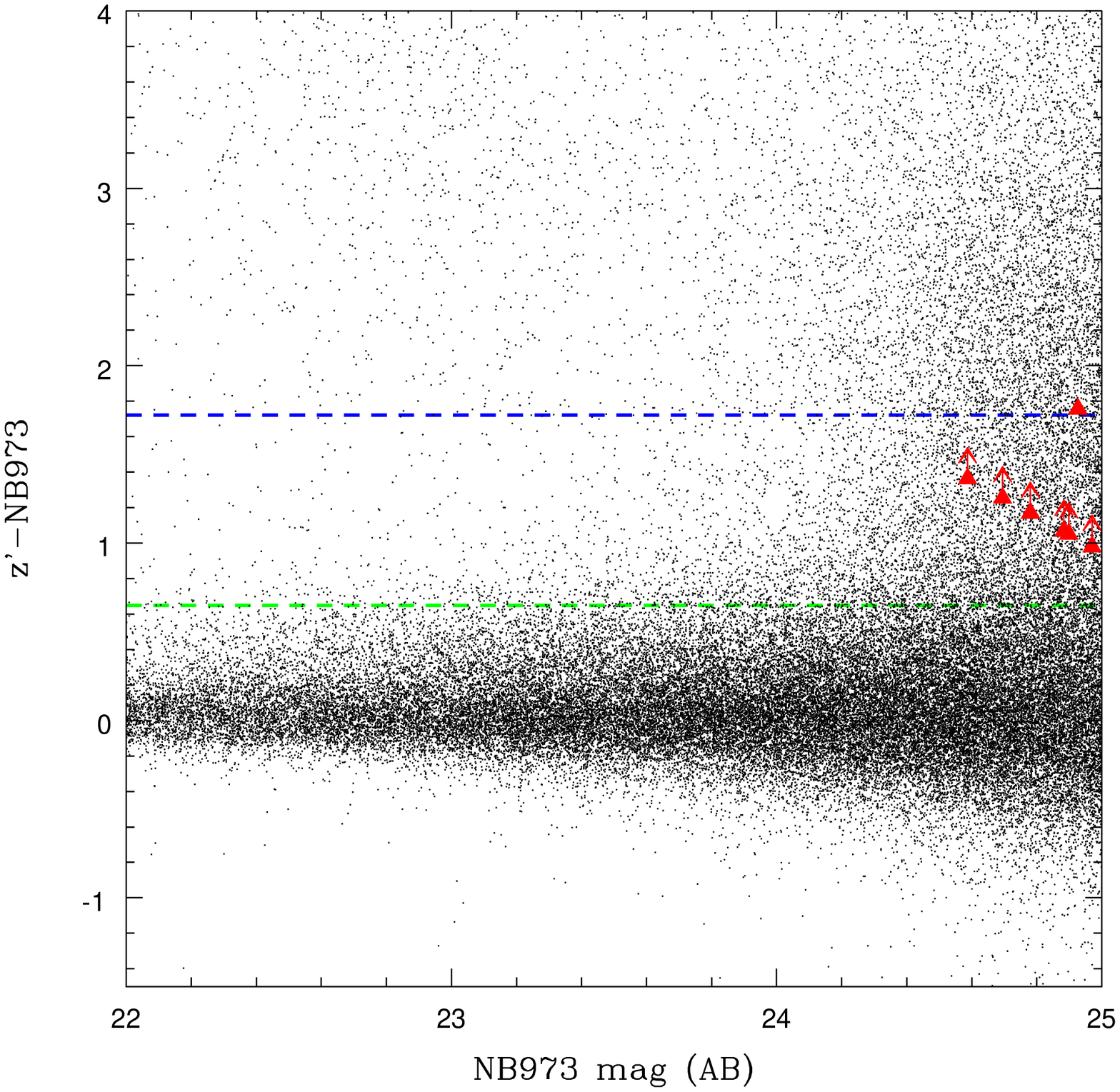}}
\caption{Color-Magnitude Diagram $z'-NB973$ vs $NB973$  for D33 field (left) and D41 (right), showing the candidates(red triangles) obtained with the criteria $z'-NB973>0.65mag$ (green dashed line). 
We also compare them to the criterion $z'-NB973>1.72mag$ (blue dashed line). The black dots represent all the objects present in each field with SNR($NB973_{combined}$)$> 5\sigma$.}
\label{fig3}
\end{figure}

\noindent{Criterion\#3 : a color criterion between $z'$ and $NB973$ data. \\}
The $NB973$ filter used for this study is included in the broad z band
filter. Although we are expecting an excess of flux in $NB973$, it is yet
possible to observe part of the continuum in the z band
filter. From Criterion\#1 and Criterion\#2, we can define the following color criterion :  $z'-NB973>0.65mag$.
We used the same NB973 filter than \cite{Ota2008} have used for the search of robust z=7 LAEs in the SDF field. Their search was successful as it resulted on the spectroscopic confirmation of the only z$\sim$6.96 spectroscopically confirmed LAE. Following the simulation they performed and plot on their Figure 3, we also check the color $z'-NB973$ of our candidates with a second color criterion : $z'-NB973>1.72mag$.
We show our candidates in the Figure~\ref{fig3} representing $z'-NB973$ vs $NB973$ for both fields, and the two color criteria. The points answering $z'-NB973>0.65mag$ but not $z'-NB973>1.72mag$ are the points with no $z'$ band detection. We used then the $3\sigma$ limit of the $z'$ band data for these objects.\\

\noindent{Criterion\#4 : $NB973-J<0$, $NB973-H<0$ and $NB973-Ks<0$.\\}
 To avoid some contamination by low-redshift objects, we choose to set NIR criteria. More explanations are given in the section \ref{contaminants}. \\

\noindent{Criterion\#5 : High redshift objects are identified as compact point-like sources.} After applying carefully all the criteria presented above, we inspect
each candidate visually to confirm their point-like source aspect. Objects with a different aspect are more likely artefacts. 7 objects out of 9 for D33 and 7 objects out of 10 for D41, have been selected as serious candidates.\\

\subsection{Contaminants} \label{contaminants}

\subsubsection {Transient objects} 

Since the narrow band and broad band data were taken at different times, it is possible for transient objects to appear in one filter and not another. Transient objects that are brighter than the narrow band detection limit will be considered as candidates if they are not visible in the broad band images.\\
\cite{Kulkarni2006} used a transient rate for SNe, including Type Ia, b,c and Type II, of $5 \times 10^4 \mathrm{Gpc}^3 \mathrm{yr}^{-1}$. From \cite{Cappellaro1999}, we know that the transient rate of Type Ia SNe is approximately a factor 3.4 lower than the transient rate including Type Ia, b, c and Type II SNe. Therefore, by applying  a transient rate of $5 \times 10^{4} / 3$ to each of our fields, we can estimate that 0.8 Type Ia SNe could be detected in the D33 field and 0.9 Type Ia SNe in the D41 field.


We chose to also estimate the number of SNe that we could have detected in our survey using the method of \cite{Ota2008}. We assume the same mean object colour than \cite{Ota2008}, $i'-NB973=0.33$, and use the results of variability. For their large magnitude variation (1.1-1.6 mag), \cite{Ota2008} find one object at each epoch giving  $P\sim10^-5$. Down to $i'\sim26.5$, we found approximately 7500 objects in D33 and 9000 objects in D41. From this method, we obtain therefore an estimated number of variable objects of $\sim$ 0.75 object for D33 and $\sim$ 0.9 object for D41 .\\

In summary, both methods seem to agree for a possible contamination of order one transient object per field.





\subsubsection{ L-T dwarfs} 

Considering Figure 9 of \cite{Hawley2002} presenting several color diagrams for L-, T- dwarfs detected in SDSS, we know that M-, L-, T- dwarfs have $z'-J > 1$ and L-, T- dwarfs have $z'-J> 2$. Therefore, although our J-band is shallower than our NB973 limiting magnitude, the criterion  $NB973-J<$0 still secure a non-contamination of our LAEs sample by L-, T-dwarfs. 
We therefore avoid selecting L- and T- dwarfs.

\subsubsection{Foreground emitters}

We estimated the lower value on equivalent width a line emitter, with a flat continuum in $f_{\nu}$, would require to be selected with our criteria using the formula from \cite{Rhoads2001}:
\begin{equation} \label{ew}
EW_{min} \sim \left(\frac{f_{NB}}{f_{BB}}\right) \Delta\lambda_{NB} = \left[\frac{5\sigma_{NB}}{3\sigma{BB}}-1\right] \Delta\lambda_{NB}
\end{equation}
with $f_{NB}$ and $f_{BB}$ the flux in $NB973$ and $g'$ band respectively, $\Delta\lambda_{NB}$ the width of the $NB973$ filter, $\sigma_{NB}$ and $\sigma_{BB}$, the flux uncertainties in $NB973$ and $g'$ band respectively.
We obtained therefore an $EW_{min}^{D33}\sim 840\AA$ and  $EW_{min}^{D41}\sim 714.2\AA$. Foreground line emitters would then require an equivalent width $EW_{min}^{D33}\ge 840\AA$ and $EW_{min}^{D41}\ge 714.2\AA$ to contaminate our \La\ selection sample.\\

1- \Ha\ at z$\sim$0.47\\

\cite{Salzer2005} presents the typical equivalent width of low redshift galaxies. They found that \Ha\ emitters at z$\sim$0.4 have $EW < 200\AA$. From Equation~\ref{ew}, the lower limit on the \Ha\ emitters equivalent width in our survey is 840\AA\ in D33 field and 714.2\AA\ in D41 field. \Ha\ emitters at z$\sim$0.47 can therefore not contaminate our candidates selection.\\

2- \Ot\ at z$\sim$0.95\\ 

\cite{Xia2010} studied the line emitters in the PEARS survey and observed that \Ot\ emitters at z$\sim$1 have generally an equivalent width lower than 500\AA .    
 Moreover, looking at the typical Spectral Energy Distributions (SEDs) of \Ot\ emitters at z$\sim$0.9, we remark that these foreground emitters are brighter in $K$s band than in $J$ band and with typical $J$ band magnitude $\sim$ 23 (AB). We will have therefore observed these objects in our very deep NIR data.
Considering the lower limit on equivalent width that line emitters will have to contaminate our candidate sample and the NIR magnitudes of typical z$\sim$0.9-1 \Ot\ emitters, the contamination by these foreground emitters is therefore ruled out.\\
\cite{Cardamone2009} found low redshift \Ot\ emitters with very high observed EWs, reaching values until 1500\AA. However, these emitters have very bright magnitudes in optical bands : $18\leq g',r',i',z' \leq 21$ and blue spectra. They would be therefore easily detected by the CFHT-LS data. These emitters are identified as Pea galaxies : luminous blue compact galaxies. \\

3- \Od\ at z$\sim$1.6\\ 

\cite{Hayashi2011} studied \Od\ emitters at z$\sim$1.46. Figure 8 of \cite{Hayashi2011} shows these emitters in a colour magnitude diagram $z'-K_{s}$ vs $K_{s}$. z$\sim$1.46 \Od\ emitters have $K_{s}\epsilon [20;23]$, and $z'-K_{s}>0$. As our narrow-band filter is included in the $z'$ band filter, as seen in Figure~\ref{fig2}, we can assume that $z'-K_{s}>0$. This colour does not agree with our Criterion\#4. 
Figure 10 of \cite{Hayashi2011} shows the colour-colour diagram $z'-K_{s}$ vs $J-K_{s}$. Most of their \Od\ emitters have not only a $z'-K_{s}>0$ but also a $J-K_{s}>0$. In the case of our sample, a deeper $J$ band image would help us eliminate the \Od\ emitters as a possible contaminant to verify that our candidates have $J-K_{s}<0$, as deduced from Criterion\#4.\\
However, our $K_{s}$ band data do not cover entirely both fields. Although the contamination by z$\sim$1.6 \Od\ emitters is very unlikely due to the $K_{s}$ magnitude range, we cannot completely rule it out.

\subsubsection{Red Continuum Galaxies}
\begin{itemize}

\item{Balmer break galaxies}

Balmer break galaxies at z$\sim$1.4 need to be also considered as a possible contamination source. However, from Figure 4 from \cite{Delorme2010}, we obtain an estimation of the colour $z' - J$ for z$\sim$2 Balmer break galaxies : $z'-J > 0$. \cite{Daddi2004} shows that z$\geq$ 1.4 Balmer Break galaxies have $z'-Ks > 1.5$. Moreover, in the case of continuum objects, we can assume that $z'-Ks \sim NB973 -Ks$.
During the selection of our candidate sample, we applied Criterion\#4 : $z'-J < 0$. If the candidates were Balmer break galaxies, we should have detected them in J with a brighter magnitude in J than in $z'$. 

\item{Extremely Red Objects (EROs)} 

We add a criterion to avoid selecting extremely red objects \citep{Cimatti2002}. We constrain this contamination by applying $NB973-J<0$, and also verifying $NB973-H<0$ and $NB973-Ks<0$.\\
\end{itemize}

These red continuum galaxies are therefore generally brighter in $Ks$ band than in NB973. These objects cannot contaminate our candidate samples. However, as our NIR data do not cover our entire fields, we cannot completely rule out this contamination.

\subsubsection{False Detections}
In order to estimate the number of false detections that could pass our selection criteria, we create an inverse $NB973$ combined image by multiplying this $NB973$ image by -1. We then applied the first criteria used for the candidate selection : SNR($NB973_{combined}$)$> 5\sigma$. None detection meets this condition. We found out therefore that our candidate samples are not contaminated by false detections.

\subsection{Summary}
After this analysis, we can conclude that it is unlikely that our z$\sim$7
LAE candidates is contaminated by low-redshift interlopers.\\
A more detailed study of the foreground emitters (\Ha\ at $\sim$0.47,
\Ot\ at $\sim$ 0.95 and \Od\ at $\sim$ 1.6) will be presented in a
forthcoming paper.\\

\subsection{Final sample}
We obtain a final sample in each observed field.
The D33 final sample contains 7 z$\sim$7 LAEs candidates over the range of $NB973$=24.3-24.8 and SNR ($NB973$)=5.2-7.7, and the D41 finale sample, 7 z$\sim$7 LAEs candidates in a range of $NB973$=24.6-25.0 and SNR ($NB973$)=5.6-7.9, as described in Table~\ref{tab:cand1} and Table~\ref{tab:cand2}. 
Using the equation given below, we estimate the lower limits of the rest-frame
equivalent widths ($EW$) derived from the photometric data and report them in Table~\ref{tab:cand1} and Table~\ref{tab:cand2}.
It is interesting to observe that the candidates from the D41 field, deeper than the D33 field (see Table~\ref{table1}) are automatically fainter than the D33 ones. The observational conditions for both fields show a systematic difference of at least 0.1 arcsec in seeing. 

\begin{eqnarray}
EW_\mathrm{rest} = \left(\frac{f_{NB973} \Delta\lambda_{z'}-f_{z'} \Delta\lambda_{NB973}} {f_{z'} - f_{NB973}} \right) \times \frac{1}{1 + z}
\label{eq:ew}
\end{eqnarray}

where $f_{NB973}$ is the observed flux in the narrow-band combined image, $f_{z'}$ is the observed flux in the $z'$ broad-band image, $\Delta\lambda_{NB973}$  and $\Delta\lambda_{z'}$ are the width of the $NB973$ filter (200\AA) and the $z'$ band filter (928\AA) respectively.  
 
Following \cite{Ota2008}, we assume that  77\% of the NB flux comes from the \La\ line : $f_{Ly\alpha}\sim 0.77 f_{NB973}$. 
We use the detection limit in $z'$ band, to derive in turn lower $EW$ limits. 

In the D33 field, LAE\#1 has a $NB973$ magnitude corresponding to 94\% completeness and the LAE\#7 $NB973$ magnitude to 89\% completeness.
In the D41 field, LAE\#1 has a $NB973$ magnitude corresponding to 96\% completeness and the LAE\#7 $NB973$ magnitude to $\sim$89\% completeness.
To estimate the completeness,  we added 200 artificial star-like objects per bin of 0.1 magnitude in blank
regions of the stacked images. We then run SExtractor on the image with the same parameters as previously used for object detection. This procedure has been repeated 20 times. The average count on 20 times of the number of artificial stars retrieved in each magnitude bin provides a direct measure of the completeness limit.\\


\begin{table*}
\caption{Table of the z$\sim6.96$ LAE candidates for the D33 field.}
\label{tab:cand1}
\centering
\begin{tabular}{c c c c c c c c} \hline \hline
\textbf{Name}   & $NB973$ & Error & SNR ($NB973$) &  $z'$ & Error & SNR ($z'$)  &$EW^{\mathrm{a}}$ (\AA)\\ \hline 
LAE\#1 & 24.3 & 0.17 & 7.7  & $>$ 25.46  & ... & ... & $>$ 24  \\        
LAE\#2 & 24.6 & 0.2  & 6.1  & $>$ 25.46  & ... & ... & $>$ 24.3  \\        
LAE\#3 & 24.7 & 0.21 & 5.7  & $>$ 25.46  & ... & ... & $>$ 24.3  \\        
LAE\#4 & 24.7 & 0.21 & 5.6  & $>$ 25.46  & ... & ... & $>$ 24.3  \\   \hline  
LAE\#5 & 24.7 & 0.22 & 5.5  & $>$ 25.46  & ... & ... & $>$ 24.3   \\   \hline  
LAE\#6 & 24.8 & 0.22 & 5.3  & 26.5 & 0.31 & 3.7 & 1   \\   \hline  
LAE\#7 & 24.8 & 0.23 & 5.2  & 26.8 & 0.38 & 2.8 & 7.6  \\   \hline  
\end{tabular}
\begin{list}{}{}
\item[$^{\mathrm{a}}$] In the rest-frame
\end{list}
\end{table*}

\begin{table*}
\caption{Table of the z$\sim6.96$ LAE candidates for the D41 field.}
\label{tab:cand2}
\centering
\begin{tabular}{c c c c c c c c} \hline \hline
\textbf{Name}   & $NB973$ & Error & SNR ($NB973$) & $z'$  & Error & $SNR (z')$  & $EW^{\mathrm{a}}$ (\AA)\\ \hline 
LAE\#1 & 24.6 & 0.17 & 7.9  & $>$ 25.4 & ...  & ... & $>$ 24.3  \\        
LAE\#2 & 24.7 & 0.18 & 7.1  & $>$ 25.4 & ...  & ... & $>$ 24.4  \\        
LAE\#3 & 24.8 & 0.19 & 6.6  & $>$ 25.4 & ...  & ... & $>$ 24.5  \\        
LAE\#4 & 24.9 & 0.2  & 6.0  & $>$ 25.4 & ...  & ... & $>$ 24.5  \\   \hline  
LAE\#5 & 24.9 & 0.3  & 5.9  & $>$ 25.4 & ...  & ... & $>$ 24.5  \\   \hline  
LAE\#6 & 24.9 & 0.21 & 5.8  & 26.7     & 0.38 & 2.8 & 4.4  \\   \hline  
LAE\#7 & 25.0 & 0.22 & 5.6  & $>$ 25.4 & ...  & ... & $>$ 18.4  \\   \hline  
\end{tabular}
\begin{list}{}{}
\item[$^{\mathrm{a}}$] In the rest-frame
\end{list}
\end{table*}

\begin{figure}
\centering
\includegraphics[width=13.cm]{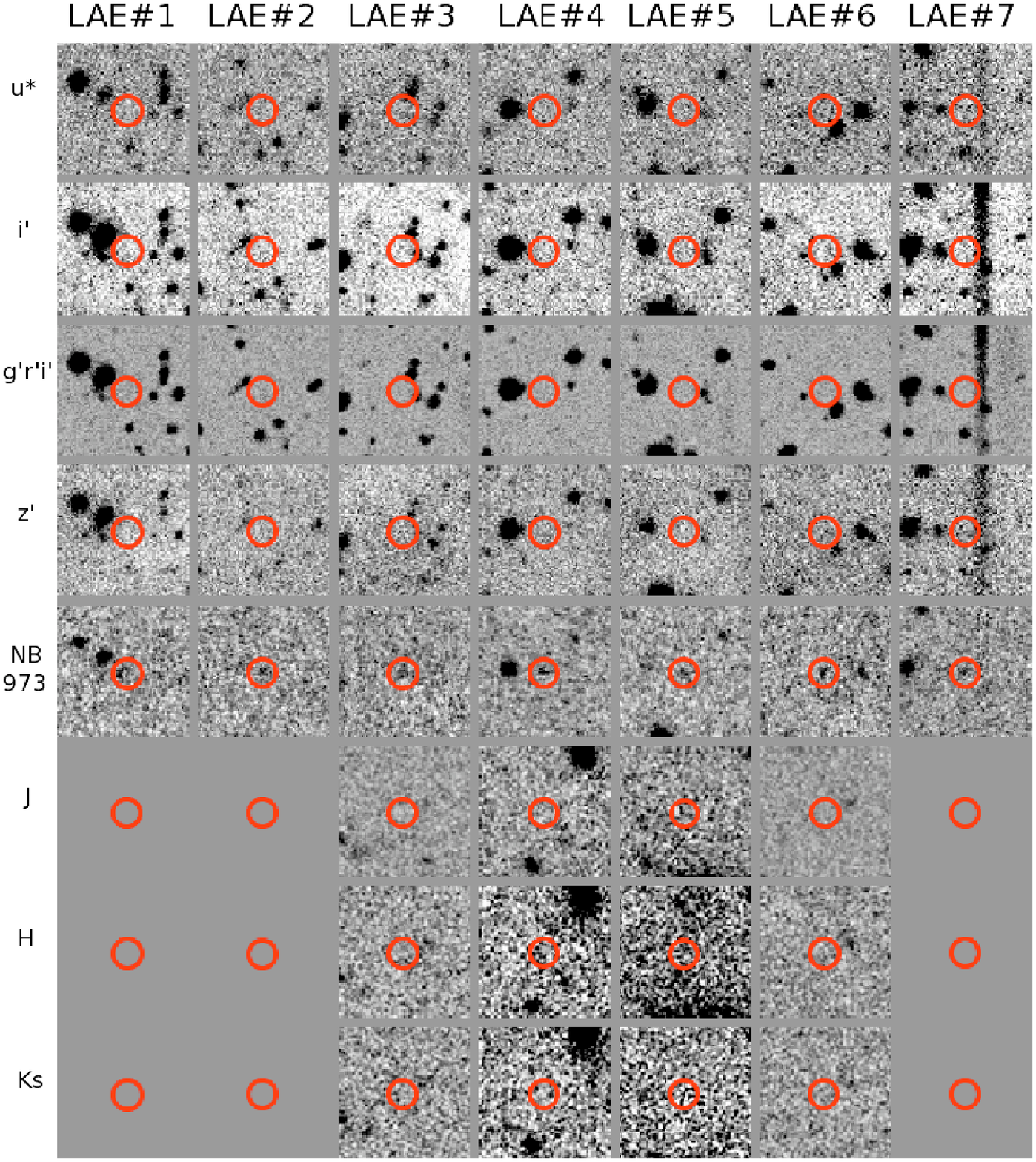}
\caption{Thumbnail images of the D33 candidates listed
in Table~\ref{tab:cand1}. Each window is $15''\times 15''$.  Objects names and passbands
are located above and to the left of the thumbnails, respectively.}
\label{fig:thumbnails33}
\end{figure}

\begin{figure}
\centering
\includegraphics[width=13.cm]{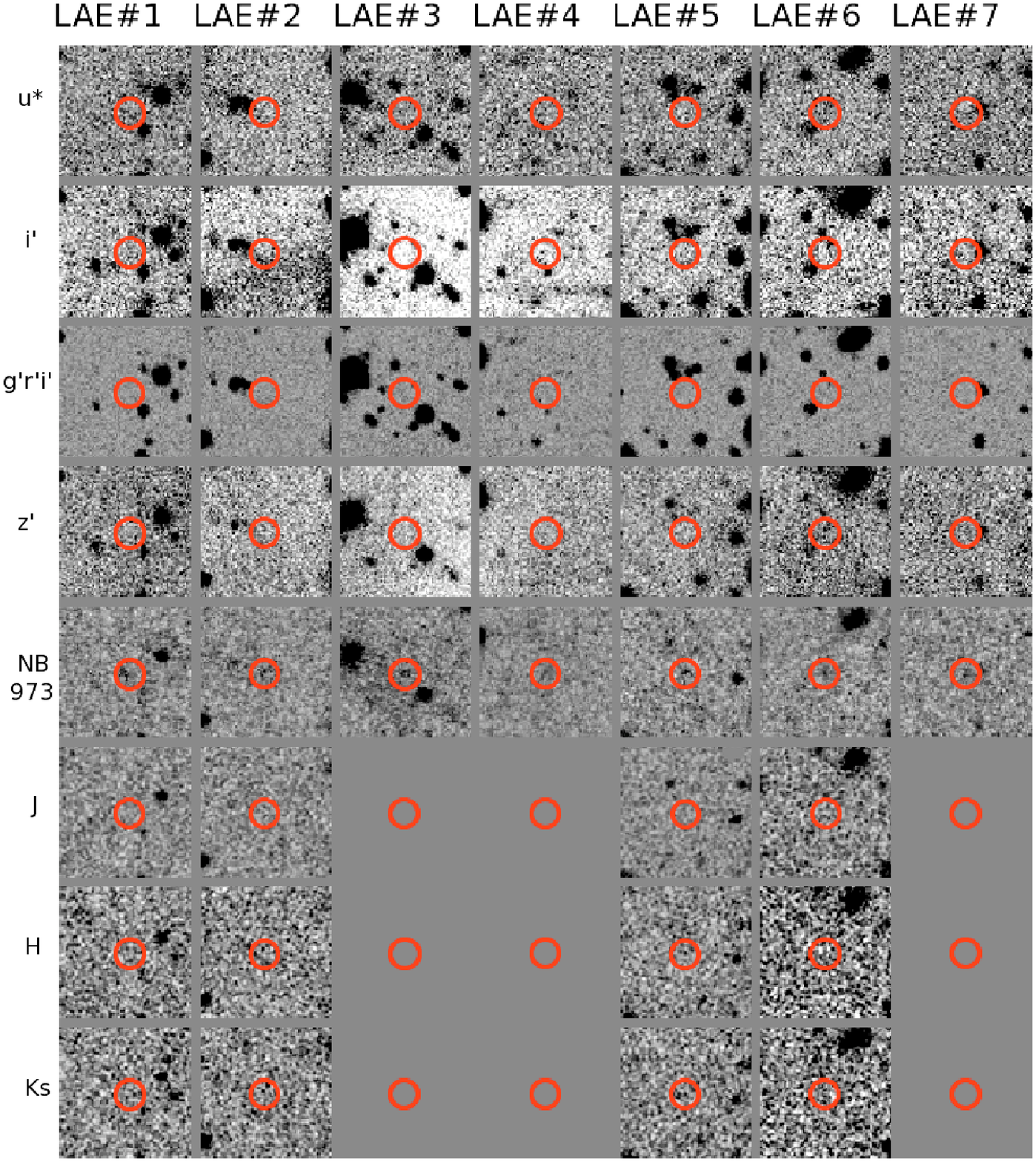}
\caption{Thumbnail images of the D41 candidates listed
in Table~\ref{tab:cand1}. Each window is $15''\times 15''$.  Objects names and passbands
are located above and to the left of the thumbnails, respectively.}
\label{fig:thumbnails41}
\end{figure}

\section{Discussion}

\subsection{Variance} \label{variance}
Two sources of variance can affect a high redshift study
: the Poisson variance and the cosmic variance, in other words the fluctuations in the large scale distribution of the galaxies. 
In order to estimate the value of this cosmic variance we used the on-line
calculator\footnote{http://casa.colorado.edu/~trenti/CosmicVariance.html}
 from the model of \cite{Trenti2008}.
We obtain then a value of 37\% and  36\% of cosmic variance in D33 and D41 fields respectively.\\
Considering the limited number of objects in our sample and the large 
comoving volume of our survey, our results are as limited by Poisson 
noise -- $\sim$38\% for the 7 objects of D33 field and for the 7 objects of D41 field -- than by clustering.
Therefore, in the case where the full sample in each observed field (7 candidates in each field) is taken into account (see Paragraph~\ref{allreal}), the error bars represent the Poisson variance and the cosmic variance. In the two other cases (see Paragraphs ~\ref{bright} and ~\ref{common}), our results are more limited by the Poisson noise than by clustering. The error bars are therefore representing the Poisson noise only.\\ 
For D33 field, we obtain a total fractional error on number counts of 0.65. As we obtain 7 objects in D33, we should expect between 4.5 and 9.4 objects in D41, if the field-to-field variation is within 1$\sigma$. As we found 7 objects D41 field, we can therefore confirm that the field-to-field variation is within 1$\sigma$. This result was expected as none of our field are known to be located in overdense or underdense large-scale structure at this redshift.
\subsection{Luminosity Function}\label{LFpart}

Following \cite{Ota2008}, we assume that on average 77\% of the narrow-band flux comes from the \La\ line. We therefore apply the same correction factor during the conversion from $NB973$ to \La\ fluxes.\\

We fit to the \La\ luminosity function
of these z$\sim$7 LAE samples, a Schechter function, $\Phi(L)$, given by
\begin{equation}
\Phi(L)\mathrm{d}L=\Phi^{*}\left(\frac{L}{L^{*}}\right)^{\alpha}\mathrm{exp}\left(-\frac{L}{L^{*}}\right)\frac{\mathrm{d}L}{L^{*}}
\end{equation}
in order to compare with previous high redshift works
\citep{Hibon2009, Ouchi2010, Ouchi2008, Ota2008}.
The error bars shown in Figure~\ref{lffull} represent the total errors (cumulative Poisson errors and cosmic variance) as explained in the previous section~\ref{variance}. In the case of Figure~\ref{lfus} and ~\ref{lfall}, the results are more limited by the Poisson noise than by the cosmic variance. The error bars represent then only the cumulative Poisson errors.
Considering the low number of candidates in our sample, we choose to
fit two out of three of the Schechter function parameters. Following
\cite{Hibon2009} and \cite{Ouchi2008}, we set the faint end slope of the
luminosity, $\alpha$, to $\alpha=-1.5$, and determine $\Phi^{*}$ and
$L^{*}$ by $\chi^{2}$ minimization.\\
We corrected the data from the completeness by number weighting.

We decided to study the following different cases :
\begin{itemize}
\item The full sample of each field is real.
\item The samples are contaminated by 50\% . We then consider only the 4 brightest objects of each field as real (referred later as bright samples). This is justified by the fact that the brightest candidates are the most robust ones and we are focusing on building the bright end of the z$\sim$7 \La\ luminosity function.
\item We chose to derive a common \La\ LF for the D33 and D41 LAEs brightest candidates. We have therefore a sample of 8 bright objects in a total area of 2320$\textrm{arcmin}^{2}$, corresponding to the sum of the D33 and D41 effective areas. 
\end{itemize}

The best-fit Schechter LF parameters for each case are summarized in Table~\ref{tab:fitlf}.

\begin{table}[!h]
\caption{Best-fit Schechter LF parameters for $\alpha = -1.5$}
\label{tab:fitlf}
\centering
\begin{tabular}{c c c} \hline \hline
Redshift & $\mathrm{log}(L^{*} (\mathrm{erg}\, \mathrm{s}^{-1}))$ &
$\mathrm{log}(\Phi^{*}(\mathrm{Mpc}^{-3}))$ \\ \hline
6.96$^{\mathrm{(1)}}$ & $42.50^{+0.1}_{-0.2}$ & $-2.40^{+0.15}_{-0.2}$  \\
6.96$^{\mathrm{(2)}}$ & $42.53^{+0.1}_{-0.2}$ & $-2.15^{+0.1}_{-0.1}$  \\ \hline
6.96$^{\mathrm{(3)}}$ & $42.73^{+0.1}_{-0.2}$ & $-3.17^{+0.15}_{-0.2}$  \\
6.96$^{\mathrm{(4)}}$ & $42.56^{+0.1}_{-0.2}$ & $-2.73^{+0.1}_{-0.1}$ \\\hline
6.96$^{\mathrm{(5)}}$ & $43.71^{+0.1}_{-0.1}$ & $-3.33^{+0.1}_{-0.1}$  \\\hline \hline
6.96$^{\mathrm{(6)}}$ & $42.8^{+0.12}_{-0.14}$ & $-3.44^{+0.20}_{-0.16}$ \\
6.5$^{\mathrm{(7)}}$ & $42.64^{+0.1}_{-0.1}$ & $-3.07^{+0.13}_{-0.13}$ \\ 
5.7$^{\mathrm{(8)}}$ & $42.8^{+0.16}_{-0.16}$ & $-3.11^{+0.29}_{-0.31}$  \\\hline
\end{tabular}
\begin{list}{}{}
\item[References.] (1, 2) derived from our D33 and D41 full samples resp.; (3, 4) derived from the brightest candidates in each field; (5) derived from a common sample between D33 and D41 fields; (6)\cite{Ota2008}; (7) \cite{Ouchi2010}; (8) \cite{Ouchi2008}
\end{list}
\end{table}

\subsubsection{All the candidates are real.} \label{allreal}

\begin{figure}
\centering
\resizebox{\hsize}{!}{\includegraphics{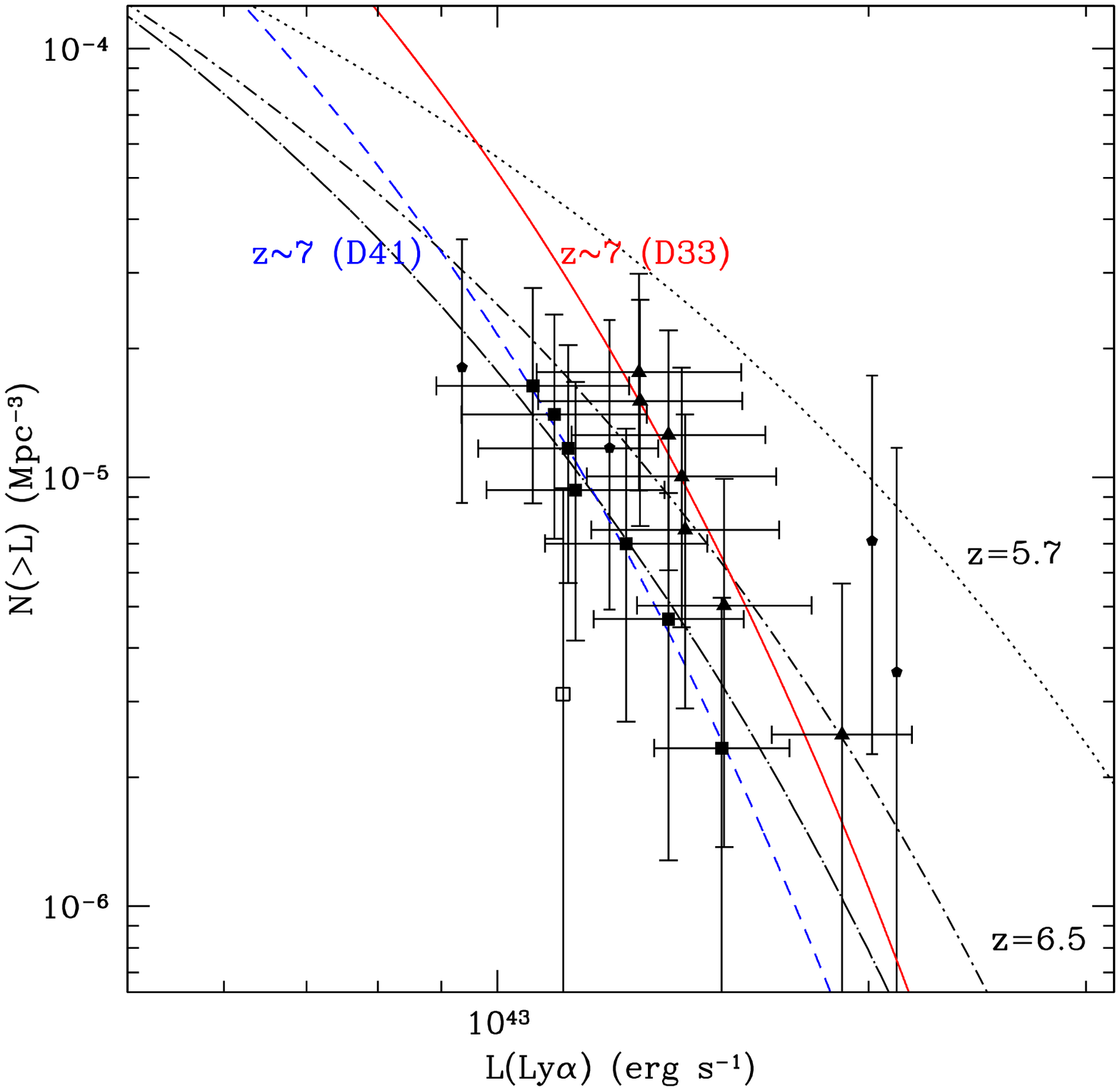}}
\caption{Best-fit Schechter function for the cumulative z$\sim$7 \La\ luminosity function derived from the D33 full sample in red and from the D41 full sample in blue. Our D33 candidates are represented as triangles and the D41 ones as filled squares, the z=6.96 LAE from \cite{Iye2006} is the empty square and \cite{Ota2008} candidates are pentagons. Also represented here is the z=6.5 \La\ LF from \cite{Kashikawa2011} (dot-short dashed line), the z=6.5 \La\ LF from \cite{Ouchi2010} (dot-long dashed line) and the z=5.7 \La\ LF from \cite{Kashikawa2011} (dotted line). The cumulative Poisson errors and the cosmic variance are taken in account in the vertical error bars. }
\label{lffull}
\end{figure}


From the z$\sim$7 \La\ LF derived from the full sample of the D41 field, seen in Figure~\ref{lffull} as the blue line and the candidates corresponding represented as filled squares, we observe a possible but weak evolution in luminosity from the z$\sim$6.5 \La\ LF from \cite{Ouchi2010} (dot-long dashed line in Figure~\ref{lffull}), but no possible evolution from the z$\sim$6.5 \La\ LF from \cite{Kashikawa2011} (dot-short dashed line in Figure~\ref{lffull}). The z$\sim$7 \La\ LF from the complete sample of the D33 field, shown as the red line and the filled triangles in Figure~\ref{lffull} shows a weaker evolution in luminosity from the z$\sim$6.5 \La\ LF.
However, looking at the error bars from D33 and D41 field candidates, both weak evolutions in luminosity, become very discussable. From these possible z$\sim$7 \La\ LFs, if all the candidates are real for one of the observed field, we cannot therefore conclude about a possible evolution in luminosity from z$\sim$6.5 to z$\sim$7.

\subsubsection{Only the brightest candidates are real.}\label{bright}

\begin{figure}
\centering
\resizebox{\hsize}{!}{\includegraphics{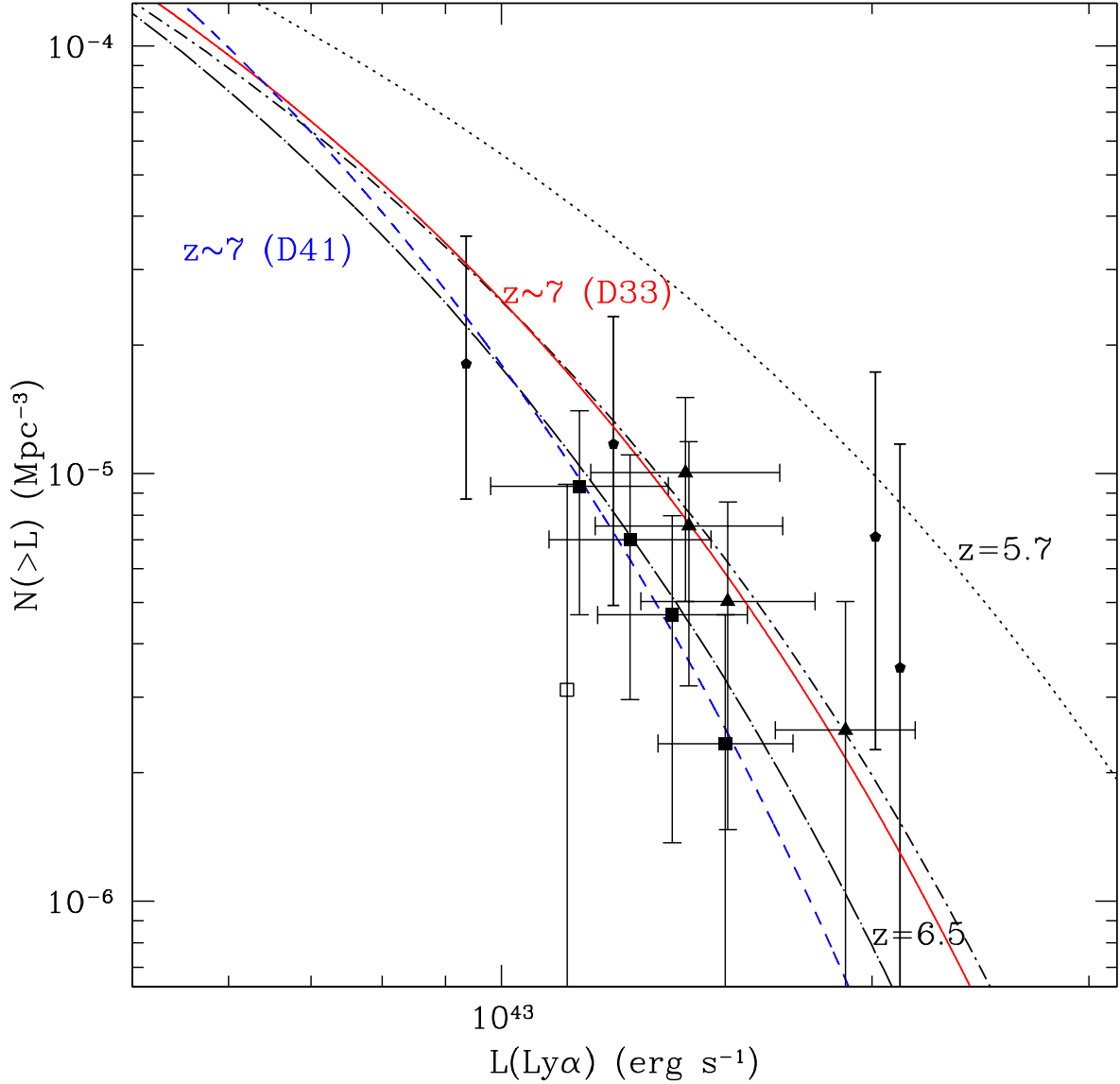}}
\caption{Best-fit Schechter function for the cumulative z$\sim$7 \La\ luminosity function derived from the D33 bright sample in red and from the D4 1 bright sample in \textbf{blue}. Our D33 candidates are represented as triangles and the D41 ones as filled squares, the z=6.96 LAE from \cite{Iye2006} is the empty square. Also represented here is the z=6.5 \La\ LF from \cite{Kashikawa2011} (dot-short-dashed line), the z=6.5 \La\ LF from \cite{Ouchi2010} (dot-long-dashed line) and the z=5.7 \La\ LF from \cite{Kashikawa2011} (dotted line). The vertical error bars represent the cumulative Poisson errors, as the bright samples are more limited by the Poisson noise than by clustering. }
\label{lfus}
\end{figure}

We derived \La\ luminosity functions for the two different z$\sim$7 bright LAE candidate samples. The best-fit Schechter functions are shown in Figure~\ref{lfus}. In both cases, we do not observe a significant evolution either in density nor in luminosity from z$\sim$6.5, from \cite{Ouchi2010} or \cite{Kashikawa2011}, to z$\sim$7.

\subsubsection{One common LF for both fields bright sample.}\label{common}

\begin{figure}
\centering
\resizebox{\hsize}{!}{\includegraphics{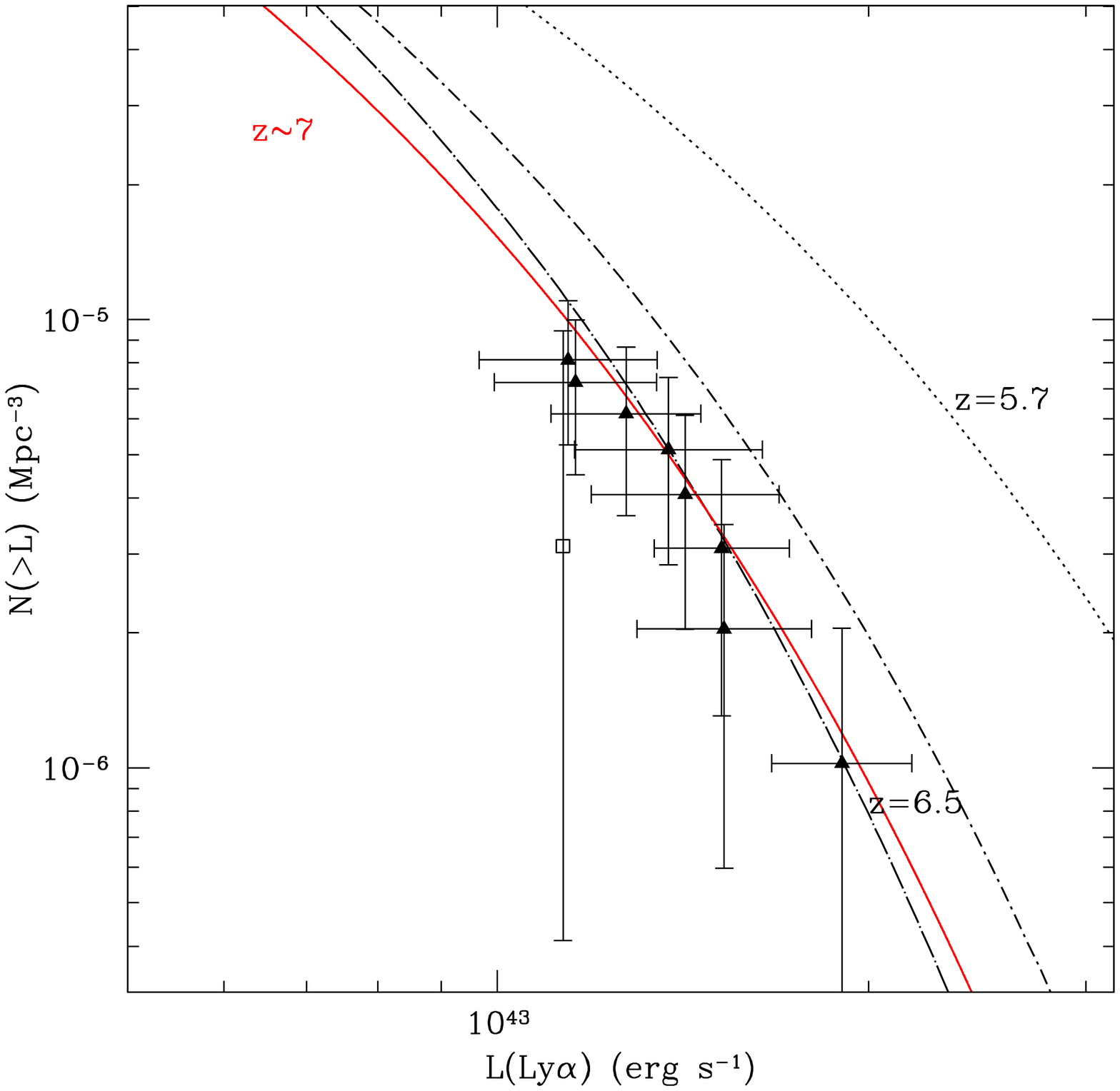}}
\caption{Best-fit Schechter function for the cumulative z$\sim$7 \La\ luminosity function derived from a common sample composed by the D33 bright sample in blue and from the D41 bright sample in red. Our D33 and D41 candidates are represented as triangles, the z=6.96 LAE from \cite{Iye2006} is the empty square. Also represented here is the z=6.5 \La\ LF from \cite{Kashikawa2011} (dot-short-dashed line), the z=6.5 \La\ LF from \cite{Ouchi2010} (dot-long-dashed line) and the z=5.7 \La\ LF from \cite{Kashikawa2011} (dotted line). The vertical error bars represent the cumulative Poisson errors, as the common sample is more limited by the Poisson noise than by clustering.  }
\label{lfall}
\end{figure}

By fitting a unique Schechter LF for both samples (as seen in Figure~\ref{lfall}), we assume that the brightest objects from both samples are real z$\sim$7 LAEs. Looking only at the filled triangles in Figure~\ref{lfall}, representing the four brightest candidates from D41 and the ones from D33, we remarked that they are fitting the z$\sim$6.5 \La\ LF from \cite{Ouchi2010}  but not the z$\sim$6.5 \La\ LF from \cite{Kashikawa2011} . Depending on the z$\sim$6.5 \La\ LF we considered, the best-fit Schechter derived LF, seen as the red line on Figure~\ref{lfall}, could agree with the observed evolution between z$\sim$5.7 and z$\sim$6.5 but would not be in favor of a strong evolution between z$\sim$6.5 and z$\sim$7.\\


\subsubsection{Summary}
We obtain two different best-fit Schechter functions for the z$\sim$7 cumulative luminosity function from our two different sets of photometric candidates. Both candidate samples help for building the bright end of the z$\sim$7 LF. Although we do not see an evolution from z$\sim$6.5 and z$\sim$7 from the D33 candidates sample, the LF derived from the D41 candidates sample do not show a significant evolution from z$\sim$6.5 to z$\sim$7, as seen in Figure~\ref{lfus}.\\
By producing z$\sim$7 LF including not only a one-field candidate sample but also the spectroscopically confirmed IOK-1, the best-fit Schechter parameters do slightly change. However our previous conclusion is still valid.\\
Assuming that the 8 brightest of our candidates are real, we interestingly obtain a z$\sim$7 LF in agreement with the z=6.5 LF produced by \cite{Ouchi2010}.\\

\cite{Ota2008} presents the first search for z$\sim$7 \La\ emitters which has lead to the first spectroscopically confirmed z$\sim$7 LAE.
 Their imaging survey covers an area of 876 square arcmin with the filter $NB973$ ($\Delta\lambda=200\AA$, $\lambda_{c}=9755\AA$) and reaches a 50\% completeness of  $NB973$=26.2 (AB, $5\sigma$). IOK-1 has a flux of $2\times10^{-17}\mathrm{erg}\, \mathrm{s}^{-1}\, \mathrm{cm}^{-2}$. 
We therefore observe a wider but shallower area. This strategy is justified by our willing to constrain only the bright-end of the z$\sim$7 \La\ LF.\\

The main prediction of \cite{LeDelliou2006} is a likely moderate decline of the bright end of the LF of LAEs from z=6.5 to z$\sim$7 arising from the evolution of the mass distribution of dark matter halos. If the LF may undergo limited evolution between z=6.5 and z=7, the effects induced by the incomplete reionization of the IGM may play an important role in the evolution of the observed LF. The faint galaxies may be more easily obscured by neutral regions and could enhance the bright end of the LF. \\
\cite{Ota2008}, by assuming that the LAEs and LBGs show a similar evolution history, can infer that the \La\ LF derived from LAEs evolves as the rest-frame UVLFs obtained from LBGs. From \cite{Yoshida2006}, they estimated therefore a possible z$\sim$7 \La\ LF with a pure luminosity evolution of  $L^{*}_{z=7}=0.58L^{*}_{z=5.7}$, with $L^{*}_{z=5.7}=1.08.10^{43}\mathrm{erg}\, \mathrm{s}^{-1}$ \citep{Shimasaku2006}. This inferred z$\sim$7 \La\ LF does not agree with our photometric candidates sample. It confirms their idea that, if LAEs are strongly related to LBGs, the neutral hydrogen fraction in the IGM is possibly higher at z$\sim$7 than at z$\sim$5.7, and this difference can cause the attenuation of \La\ lines of high redshift LAEs.\\

If none of our object is a real z$\sim$7 LAE, we can then put an upper limit on the z$\sim$7 \La\ LF, which helps for constraining better these models.


\section{Conclusions}
We observed 0.64 square degree of the WIRDS/CFHT-LS fields with the $NB973$ filter,  
corresponding to targeting the \La\ line at z$\sim$7. After applying our selection criteria and verifying that our selection was not contaminated by low-redshift emitters, we obtained a sample of seven z$\sim$7 LAEs in each field. From these photometric samples, we have been able to infer  possible z$\sim$7 \La\ Luminosity Functions. We did not find a significant evolution either in luminosity nor in density from z=6.5 to z$\sim$7.\\
It is now crucial to obtain spectroscopic follow-up observations to reveal the real nature of these objects and establish a firm conclusion on the z$\sim$7 \La\ Luminosity Function.
 The exact evolution of the LF beyond redshift 6.5 remains therefore a matter to debate. Moreover, a single galaxy at z$\sim$7 is obviously not sufficient to firmly constrain the bright end of the Ly$\alpha$ LF. We need to increase the number of independent fields as well as the number of LAEs at z=7.
Once the bright end of z$\sim$7 Ly$\alpha$ LF is determined and possible evolution from z=6.5 is derived, it will become easier to assess whether the 1.06$\mu$m and the 1.19$\mu$m NB filters could reveal z$\sim$8-9 LAEs.\\

\bibliographystyle{apj}
\bibliography{SCAM_v6}
\end{document}